%% file: conference_101719.tex
\documentclass[conference]{IEEEtran}
\IEEEoverridecommandlockouts

\input{macro}
\def\BibTeX{{\rm B\kern-.05em{\sc i\kern-.025em b}\kern-.08em
    T\kern-.1667em\lower.7ex\hbox{E}\kern-.125emX}}
\begin{document}

\title{A Dense Tensor Accelerator with Data Exchange Mesh for DNN and Vision Workloads}
\author{
\IEEEauthorblockN{Yu-Sheng Lin and Wei-Chao Chen}\IEEEauthorblockA{\textit{\qquad Inventec Corporation\qquad}\\ lin.john@inventec.com, chen.wei-chao@inventec.com}
\and
\IEEEauthorblockN{Chia-Lin Yang and Shao-Yi Chien}\IEEEauthorblockA{\textit{\qquad National Taiwan University\qquad}\\ yangc@csie.ntu.edu.tw, sychien@media.ee.ntu.edu.tw}
}
\maketitle

\input{00_abstract}
\input{10_intro}
\input{30_architecture}
\input{40_experiment}
\input{50_conclusion}

\pagebreak
\balance
\bibliographystyle{unsrt} 
\bibliography{refs}

\end{document}

%% file: macro.tex
\usepackage{mathptmx}
\usepackage{balance}
\usepackage{fancyhdr}
\usepackage[normalem]{ulem}
\usepackage[hyphens]{url}
\usepackage[sort,nocompress]{cite}
\usepackage[final]{microtype}
\usepackage[keeplastbox]{flushend}
\usepackage[bookmarks=true,breaklinks=true,letterpaper=true,colorlinks,linkcolor=black,citecolor=blue,urlcolor=black]{hyperref}
\usepackage{cite}
\usepackage{amsmath,amssymb,amsfonts,array}
\usepackage{algorithm,algorithmic}

\floatname{algorithm}{Procedure}
\usepackage{graphicx}
\usepackage{verbatim}
\usepackage{subfig}
\usepackage{textcomp}
\usepackage{booktabs}
\usepackage[dvipsnames, svgnames, x11names]{xcolor}
\usepackage{soul,tabu}
\definecolor{lightyellow}{RGB}{255,255,160}
\sethlcolor{lightyellow}
\usepackage{siunitx}
\newcommand{\ArchName}{VectorMesh}
\newcommand{\myparagraph}{\vspace{\baselineskip}\noindent\textbf}
\newcommand{\figrtext}{Fig.}
\newcommand{\tabrtext}{Table}

\newcommand{\secrtext}{Section}
\newcommand{\eqrtext}{Eq.}
\newcommand{\figref}[1]{\figrtext~\ref{fig:#1}}

\newcommand{\tabref}[1]{\tabrtext~\ref{tab:#1}}

\newcommand{\equref}[1]{\eqrtext~(\ref{equ:#1})}

\newcommand{\secref}[1]{\secrtext~\ref{sec:#1}}

\newcommand{\etal}{\textit{et~al}.}
\newcommand{\ie}{\textit{i}.\textit{e}.,}

\newcommand{\bA}{\mathbf{A}}
\newcommand{\bB}{\mathbf{B}}
\newcommand{\bC}{\mathbf{C}}

\newcommand{\bE}{\mathbf{E}}
\newcommand{\bF}{\mathbf{F}}
\newcommand{\bG}{\mathbf{G}}
\newcommand{\bH}{\mathbf{H}}
\newcommand{\bI}{\mathbf{I}}

\newcommand{\bP}{\mathbf{P}}
\newcommand{\bQ}{\mathbf{Q}}
\newcommand{\bR}{\mathbf{R}}
\newcommand{\bS}{\mathbf{S}}

\newcommand{\bW}{\mathbf{W}}
\newcommand{\bX}{\mathbf{X}}
\newcommand{\bY}{\mathbf{Y}}
\newcommand{\bZ}{\mathbf{Z}}

\newcommand{\bk}{\mathbf{k}}

\newcommand{\bzero}{\mathbf{0}}
\definecolor{pid}{HTML}{00A0B3}
\definecolor{tid}{HTML}{F54291}

\graphicspath{{figure/}}

%% file: 00_abstract.tex
\begin{abstract} 
We propose a dense tensor accelerator called VectorMesh, a scalable, memory-efficient architecture that can support a wide variety of DNN and computer vision workloads.  Its building block is a tile execution unit~(TEU), which includes dozens of processing elements~(PEs) and SRAM buffers connected through a butterfly network.  A mesh of FIFOs between the TEUs facilitates data exchange between tiles and promote local data to global visibility.  Our design performs better according to the roofline model for CNN, GEMM, and spatial matching algorithms compared to state-of-the-art architectures. It can reduce global buffer and DRAM fetches by 2-22 times and up to 5 times, respectively.
\end{abstract}


%% file: 10_intro.tex
\section{Introduction}\label{sec:intro}
The goal of designing a fast deep neural network (DNN) accelerator involves packing as many processing elements (PEs) as possible and run them without stalling with a smooth and timely supply of data.  While the density of PEs increases with the advancement of technology, the available DRAM bandwidth tends to grow slower than computation~\cite{dram_power}.  For effective use of this precious DRAM bandwidth, some modern accelerators exploit the sparsity in DNN to reduce DRAM bandwidth~\cite{scnn,cnvlutin,gratetile}. On the other hand, dense accelerators employ on-chip buffers that can occupy more than half of the chip area~\cite{tpu,eyeriss}. Together, these resources define a performance roofline~\cite{roofline}, and an architecture can not perform at this limit if it fetches the same data from DRAM multiple times and duplicate them in the buffers.

For this purpose, modern architectures divide a workload into individual small groups so that PE can fetch the same data repeatedly from the on-chip buffers without wasting global DRAM bandwidth. This technique is called tiling, which largely determines the characteristics of each architecture.  For example, Google's Tensor Processing Unit~(TPU)~\cite{tpu,systolic} utilizes a global tiling buffer with mesh-connected PEs to create a highly scalable architecture. However, this architecture requires synchronized PE execution, resulting in bubbles when running smaller tiles in larger TPUs.  A more recent TPU instead adopts a much smaller tile size~\cite{tpu2} to alleviate this issue.
Eyeriss~\cite{eyeriss,eyerissv2} is another architecture with a smaller tile size, and each PE has a dedicated local tiling buffer for computing elements of convolution partial sum~(PSum) without accessing the global buffer. It employs a horizontal multicast network to deliver data to multiple local buffers within the same cycle.  Because each local buffer is private to its PE, data needs to be duplicated across several local buffers and the global buffer, wasting precious on-chip storage.

In conclusion, small tiles are good for keeping the PEs busy but bad for local buffer pressure due to data duplication.  The proposed \ArchName{} architecture aims to balance this tradeoff by allowing data exchange across small tiles without duplication (\figref{cover:arch}).  Its basic computing block is the Tile Execution Unit (TEU), a small tile processor with synchronized PEs. Neighboring TEUs are joined together with \textit{bidirectional FIFOs} to form a 2D mesh arrangement. \ArchName{} supports classic CNN layers~\cite{alexnet,tinyyolov1,srcnn} and variant CNN layers~\cite{pixel_shuf,dilconv}, as well as spatial matching for video inference acceleration~\cite{flownet,eva}. A TEU can support a tiled version of layer workloads through its two 32-bank local input buffers, 32 vectorized PEs~(\textit{PE group, PEG}), and a PSum buffer. The interconnect from the buffers to the PEG is a \textit{butterfly network (BFN)}, and we adopt a systematical approach to guarantee BFNs can provide full throughput for our target applications. As the throughput of BFN is guaranteed, all TEUs can run at a similar speed, and each FIFO only utilizes four entries of 32 words to balance the skew.  Our contributions include:
\begin{itemize}
    \item A dense tensor accelerator that utilizes mesh FIFOs for data exchange for DNN and computer vision workloads, 
    \item A methodology for scheduling workloads onto the proposed architecture, and
    \item An implementation in both SystemC and Verilog with competitive performance using the roofline analysis.
\end{itemize}

%% file: 30_architecture.tex
\section{The \ArchName{} Architecture}\label{sec:architecture}
The block diagram of \ArchName{} is static and incomplete without a discussion of workload scheduling.  For this, we convert the workloads into tensors (\secref{architecture:workload}), divide these tensors into tiles (\secref{architecture:tile}), share the data opportunistically using the FIFOs, and create a methodology to provide full buffer throughput to the PEs through BFNs (\secref{architecture:butterfly}).

\begin{figure*}[ht]
    \centering
    \includegraphics[width=0.85\textwidth]{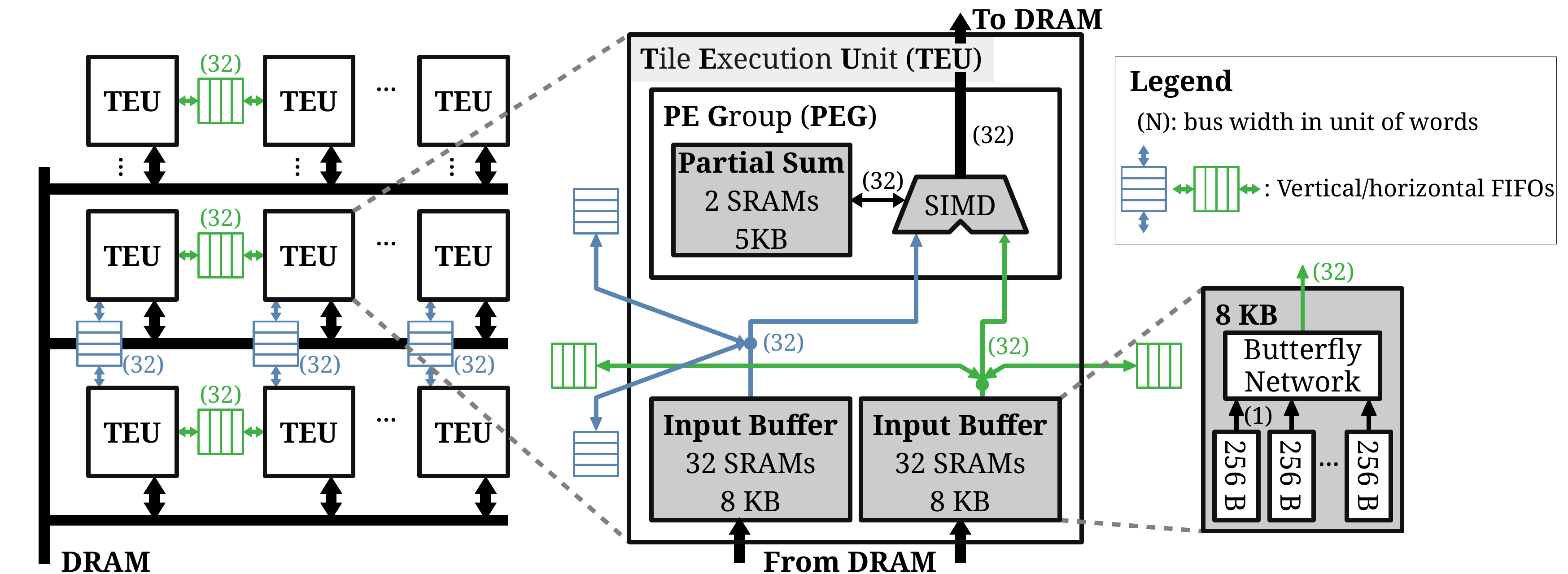}
    \caption{\bf The \ArchName{} architecture. \normalfont The architecture can execute the workloads such as DNN, CNN, and spatial matching. Using classic butterfly networks for data routing results in a simple yet efficient design.}\label{fig:cover:arch}
\end{figure*}

\subsection{Target Workloads Formulation}\label{sec:architecture:workload}
\myparagraph{Matrix Multiplication~(MM).}
\equref{architecture:gemm} shows a MM operation $\bC=\bA\bB$ in a tensor form, which involves two matrices of size $O(n^2)$ and requires $O(n^3)$ operations. Each element in the 2D matrix performs a 1D dot-product, and each dot-product can be carried out in parallel. In \equref{architecture:gemm}, they are represented as two \textit{parallel} indices $(i, j)$ in the left-hand side expression and one \textit{temporal} index $(k)$ in the summation subscript.

\begin{equation}\label{equ:architecture:gemm}\begin{aligned}
&\bC(i,j) = \sum_{k}\bR_\bA(i,j,k)\bR_\bB(i,j,k),\\
&\text{where}~\forall(i,j,k)\in\text{NDRange}(M,N,K)\,,\\
&\left\lbrace\begin{aligned}
    \bR_\bA(i,j,k) &= \bA(i,k)\\
    \bR_\bB(i,j,k) &= \bB(k,j)
\end{aligned}\right.\,.
\end{aligned}\end{equation}
\vspace*{-1em}
\myparagraph{Convolution Neural Network~(CNN).}
A CNN layer produces an output feature tensor $\bC$ ($C_o\times o_w\times o_h$) using a kernel tensor $\bk$ ($C_o\times C_i\times k_w\times k_h$) and an input feature tensor $\bI$ ($C_i\times i_h\times i_w$).
Each output element requires a $C_i\times k_w\times k_h$ convolution, which can also be written in a similar form:
\begin{equation}\label{equ:architecture:cnn}\begin{aligned}
&\bC(i,j,k) = \sum_{l,m,n}\bR_\bI(i,j,k,l,m,n)\bR_\bk(i,j,k,l,m,n),\\
&\text{where}~\forall(i,j,k,l,m,n)\in\text{NDRange}(C_o,o_w,o_h,C_i,k_w,k_h)\,,\\
&\left\lbrace\begin{aligned}
    \bR_\bI(i,j,k,l,m,n) &= \bI(l,j+m,k+n)\\
    \bR_\bk(i,j,k,l,m,n) &= \bk(i,l,m,n)
\end{aligned}\right.\,.
\end{aligned}\end{equation}
\vspace*{-1em}
\myparagraph{Spatial Matching Algorithms.}
Special matching algorithms~\cite{flownet,costvol,eva} have huge potentials when combined with modern CNNs. These algorithms require two input feature maps, namely \textit{current} and \textit{reference}. For each pixel~(block) in the \textit{current} feature map, we compute the dot-product between the \textit{current} pixel and its nearby pixels in the \textit{reference} feature map.
For example, the correlation layer~\cite{flownet} computes the spatial correlation between two tensors $\bI$ ($C_i\times o_h \times o_w$), which we can write as:
\begin{equation}\label{equ:architecture:corr}\begin{aligned}
&\bC(i,j,k,l) = \sum_{m}\bR_\mathbf{I1}(i,j,k,l,m)\bR_\mathbf{I2}(i,j,k,l,m)\,,\\
&\text{where}~\forall(i,j,k,l,m)\in\text{NDRange}(s_w,s_h,o_w,o_h,C_i)\,,\\
&\left\lbrace\begin{aligned}
    \bR_\mathbf{I1}(i,j,k,l,m) &= \bI_1(m,i,j)\\
    \bR_\mathbf{I2}(i,j,k,l,m) &= \bI_2(m,i+k,j+l)
\end{aligned}\right.\,.
\end{aligned}\end{equation}
Spatial matching algorithms require different datapath designs and therefore cannot be efficiently supported by CNN or MM processors~\cite{systolic_bm1,systolic_bm2}.

\subsection{Tiled Execution for Target Workloads}\label{sec:architecture:tile}
To make \ArchName{} adaptive to these target workloads, it must allow various permutations and combinations of the \textit{parallel} and \textit{temporal} indices in any order. This section illustrates a methodology to convert the mathematical forms above into a feasible workload scheduling in \ArchName{}.

A \ArchName{} TEU has 16~KB input buffers and a 5~KB PSum buffer available for tiling.  To obtain a valid tiling scheme, we must divide workloads into groups that fit into the buffers.  Take the MM workload~(\equref{architecture:gemm}) as an example.  A natural tiling is to divide the NDRange into rectangular groups of size $(t_i,t_j,t_k)$:
\begin{equation}\label{equ:architecture:gemm_tile}\begin{aligned}
&\bC(i,j) = \sum_{k}\bR_\bA(i,j,k)\bR_\bB(i,j,k),\\
&\text{where}~\forall(i,j,k)\in\text{NDRange}(t_i,t_j,t_k)\,.
\end{aligned}\end{equation}

Based on tensor analysis methodologies~\cite{umi,halide13}, this results in $t_it_jt_k$ MAC operations on $t_it_j$ PSums and $(t_i+t_j)t_k$ input buffers. We keep PSums with the same \textit{parallel} indices $(i,j)$ static in a TEU, such that the PSum does not consumes external bandwidth, and one MAC operation consumes ${(t_i+t_j)t_k}/{(t_it_jt_k)}$ bandwidth on average. This methodology also applies to other target workloads, and we can manually choose a valid tile size that minimizes the bandwidth for every target workloads.

This scheduling results in only one external memory write for each PSum, which is the optimal bandwidth. On the other hand, the input buffer size limits the available tile size. To overcome the limitation, we add the FIFO to share the input buffer across TEUs.

\begin{figure}[ht]
    \centering
    \subfloat[Tiles layout in phase 0.]{\includegraphics[scale=0.43,page=1]{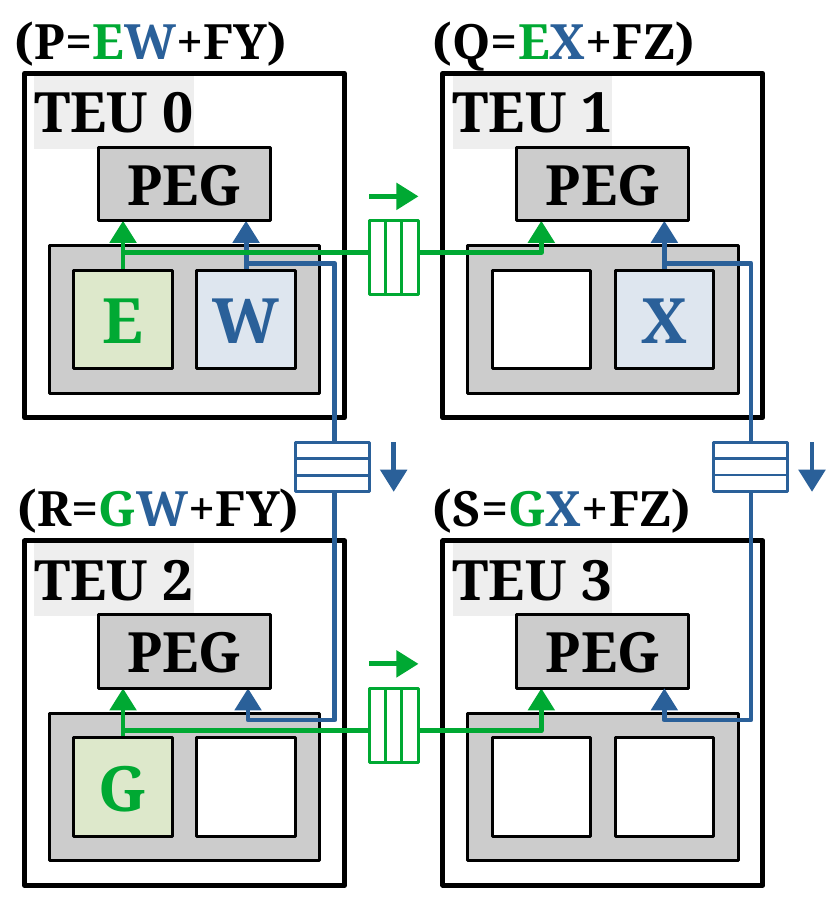}\label{fig:architecture:sd_reason:1}}\unskip\ \vrule\ \hspace{5pt}
    \subfloat[Tiles layout in phase 1.]{\includegraphics[scale=0.43,page=2]{systolic_tile.pdf}\label{fig:architecture:sd_reason:2}}%
    \caption{\normalfont The data sharing mechanism ensure all input buffer store unique data, maximizing the utilization of SRAMs.}\label{fig:architecture:sd_reason}
\end{figure}

Consider executing this GEMM example on \ArchName{}:
\begin{equation}\label{equ:architeture:gemmb22}
\underbrace{\begin{bmatrix}\bP&\bQ\\ \bR&\bS\end{bmatrix}}_\bC =
\underbrace{\begin{bmatrix}\bE&\bF\\ \bG&\bH\end{bmatrix}}_\bA ~
\underbrace{\begin{bmatrix}\bW&\bX\\ \bY&\bZ\end{bmatrix}}_\bB.
\end{equation}
\figref{architecture:sd_reason} shows a possible scheduling for sharing the input buffer. We need four TEUs to compute the output $(\bP,\bQ,\bR,\bS)$. In \figref{architecture:sd_reason:1}, the PSums $\bP = \bE\bW$ and $\bQ=\bE\bX$ would both request $\bE$, which can be shared through the horizontal FIFOs under this scheduling. To identify all shareable tensors, we first notice that the upper two TEUs process tiles with different \textit{parallel} indices $j$. Since \emph{the partial derivative of the right-hand side arguments of $\bA$ against $j$ is zero in} \equref{architecture:gemm} (\ie~$\partial(i,k)/\partial j = \bzero$), the upper two TEUs only need to read $\bE$ once.

\subsection{Executing a Tile on a TEU}\label{sec:architecture:butterfly}
A TEU has 32 vectorized PEs and can consume 32 \textit{parallel} indices from a tile per cycle. Using the same MM representation, we have:

\begin{equation}\label{equ:architecture:gemm_warp}\begin{aligned}
&\bC(i,j) = \bC(i,j)+\bR_\bA(i,j,k)\bR_\bB(i,j,k)\,,\\
&\forall(i,j)\in\text{NDRange}(p_i,p_j)\,,\quad p_ip_j=32.
\end{aligned}\end{equation}

To perform this operation, PEs must read two vectors from the input buffers through the BFN~\cite{meritz,maeri}. However, if some TEUs cannot access its vectors in one cycle, then stall occurs and propagates to all TEUs.  To prevent this from happening, we adopt a strategy as follows.  For a memory system consisting of $2^X$ banked SRAM ($X=5$ in \ArchName{}), Lin~\etal~\cite{meritz} shows that if the accessing address $A_N$ of the $N$-th PE can be written as $A_N = A_0 + \sum_{i=0}^{X} 2^Xo_i b_i$, where $b_i$ is the $i$-th digits of the binary notation of $N$, and $o_i$'s are odd numbers, then a BFN can always serve the data for this system in one cycle. Applying the padding and shuffling techniques to the local buffer data~\cite{gg3,meritz}, we ensure all our target workloads can fulfill this condition.

%% file: 40_experiment.tex
\section{Experiments}\label{sec:exp}

\subsection{Workloads Supported by \ArchName{}}\label{sec:exp:workload}
\ArchName{} supports a broader range of workloads compared with classic CNN accelerators like TPU or Eyeriss. In this section, we first benchmark the performance for \textit{typical DNN workloads} across different architectures.  Next, we show that \ArchName{} can achieve high performance for \textit{modern CNN workloads and spatial matching workloads} that do not execute efficiently on other architectures.  The following paragraphs describe these two types of workloads.

\myparagraph{Typical DNN Workloads.} We select representative DNN layers from AlexNet, TinyYOLO, Inception, and SRCNN~\cite{alexnet,tinyyolov1,inceptionv4,srcnn}. (We abbreviate them to AL, TY, IN, and SR, respectively.) As shown in \tabref{exp:standard_cnn}, these workloads cover both square and non-square kernels with sizes $(1,3\cdots,11)$.

\input{large_float/table_networks}

\myparagraph{Modern CNN Workloads and Spatial Matching Workloads.} We select the representative DNN layers from more recent networks including the DeepLab~\cite{deeplabv3}, ESPCN~\cite{pixel_shuf}, and MobileNet~\cite{mobilenet}. Also, we select spatial matching workloads~\cite{eva,flownet} discussed in \secref{architecture:workload}.

\subsection{Architectural Implementation}\label{sec:exp:impl}

We develop a cycle-level simulator to evaluate the effectiveness of the proposed VectorMesh architecture. We also implementation efficient TPU and Eyeriss simulator by tiling and prefetching. For example, the normalized performance of AlexNet on our 128-PE Eyeriss only differs slightly (10\%) from the reference implementation~\cite{eyeriss}.

The detailed simulation configuration is explained below:
\begin{itemize}
\item \myparagraph{PE Numbers and Frequency.} We simulate $N_{\text PE}=128$ and $512$ PEs version for TPU, Eyeriss, and \ArchName{}. TPU and Eyeriss are shaped as $8\times 16$, $16\times 32$ PEs, and \ArchName{} is shaped as $2\times 2$ and $4\times 4$ TEUs. The simulation also assumes a working frequency of 200~MHz.
\item \myparagraph{Bandwidth and Buffer Sizes.} We adopt a DDR simulator~\cite{ramulator} to more accurately evaluate the DRAM and global buffer subsystem. We adopts fixed 6.4GB/s DRAM bandwidth (two DDR4-1600 x16 devices) and 25.6~GB/s global buffer bandwidth. For every PE in TPU, Eyeriss, and \ArchName{}, we allocate $0,0.3,0.6$ KB local buffers and $1.0N_{\text PE},0.5N_{\text PE},2$ KB global buffers, respectively. We choose these numbers to match the PE-to-memory ratio from existing publications~\cite{tpu,eyeriss}. Also, the required size of global buffer of \ArchName{} does not need to grow with $N_{\text PE}$ accordingly.
\end{itemize}

For the circuit-level comparison, we implement \ArchName{} with Verilog. As a result of the architecture's simplicity, our codes are relatively lightweight at 9.6k lines, using 1.1M gates per TEU.
Based on our synthesizing results and the statistics from Eyeriss and TPU, we also estimate the area of different architecture given the buffer configuration above, as shown in \tabref{exp:area_norm}.

\begin{table}[ht]
\centering
\small
\caption{Area estimation of the architectures.}\label{tab:exp:area_norm}
\begin{tabular}{lrrr}
\toprule
                & Eyeriss & TPU & VectorMesh\\
\midrule
MAC             & 0.08 & 0.08 & 0.08\\
Global buffer   & 0.19 & 0.38 & 0.00\\
Local buffer    & 0.48 & 0.00 & 0.67\\
Controllers     & 0.25 & 0.00 & 0.25\\
BFN+FIFO        & 0.00 & 0.00 & 0.04\\
\hline
Area factor ($A$) & 1.00 & 0.46 & 1.04\\
\bottomrule
\end{tabular}
\end{table}

\subsection{Architectural Simulation}

\begin{figure}
\centering
\includegraphics[width=0.45\textwidth,page=1]{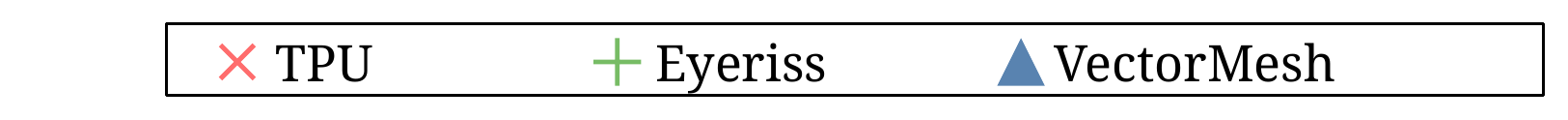}\\%
\includegraphics[width=0.45\textwidth]{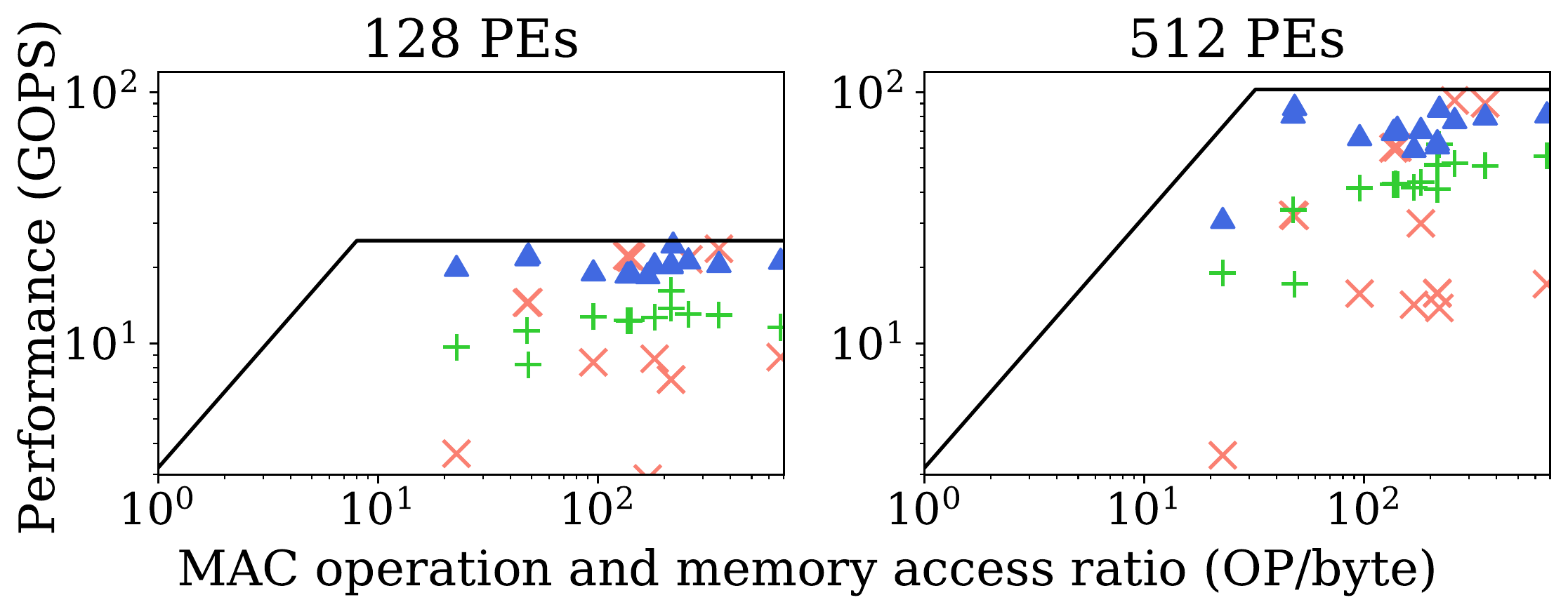}
\caption{\bf Architectural comparisons using the roofline analysis against the workloads in \tabref{exp:standard_cnn}.}\label{fig:exp:standard_workload}
\end{figure}

Based on the simulator and the hardware implementation, we discuss how different architectures can fully utilize their computation resources efficiency from three aspects:

\myparagraph{Roofline Analysis.} The roofline analysis provides an upper-bound under the same bandwidth and PE resources for different workloads. Given an infinite size and infinite bandwidth on-chip buffer, the performance upper-bound of a workload is the minimum between (1) the PE processing rate over the total MAC operations and (2) the DRAM bandwidth over total input and output data sizes.
In \figref{exp:standard_workload}, we denote the roofline as the black line. As can be seen, \ArchName{} performs closer to the roofline than others when given the same resources because it is more bandwidth efficient, as discussed in the next paragraph.

\myparagraph{Memory Access.} Lower memory access indicates higher utilization and lower power. We define the \emph{normalized access} as the number of bytes accessed from memory per 1,000 MAC operations. We calculate this for both global buffers~(GLB) and the DRAM. As shown in \tabref{exp:perf_norm}, in terms of global buffer bandwidth, TPU generates $18$-$22$x larger bandwidth than VectorMesh due to the lack of local buffer; \ArchName{} consumes $2$-$4$x less bandwidth than Eyeriss since it does not duplicate data in local buffers. As a result, even with $64$-$256$x smaller global buffers than TPU and Eyeriss, our DRAM bandwidth result is still competitive, with $-14$-$+44$\% and $2$-$5$x bandwidth reduction compared with Eyeriss and TPU, respectively.

\myparagraph{Area Efficiency.} Since different architectures require different hardware resources per PEs, for a fair comparison, it is crucial to take area efficiency into account as it reflects how much performance is available per chip area. We define this metric by dividing its average performance $P$ on all workloads by an area factor $A$ based on the estimation in \tabref{exp:area_norm}. As shown in \tabref{exp:perf_norm}, while Eyeriss and \ArchName{} provide better raw performance, only \ArchName{} provides higher area efficiency that TPU since Eyeriss suffers from data duplication across its local buffers. Also, a 128-PE TPU provides a higher area efficiency than a 128-PE \ArchName{} because when the number of PEs is small, both architectures utilize small tiles, which match the workload.  The routing simplicity of TPU thus produces a marginally better efficiency.

\input{large_float/table_all}

\subsection{Workloads Adaptiveness of \ArchName{}}

\begin{figure}
\centering
\includegraphics[width=0.45\textwidth,page=2]{legend.pdf}\\%
\includegraphics[width=0.45\textwidth]{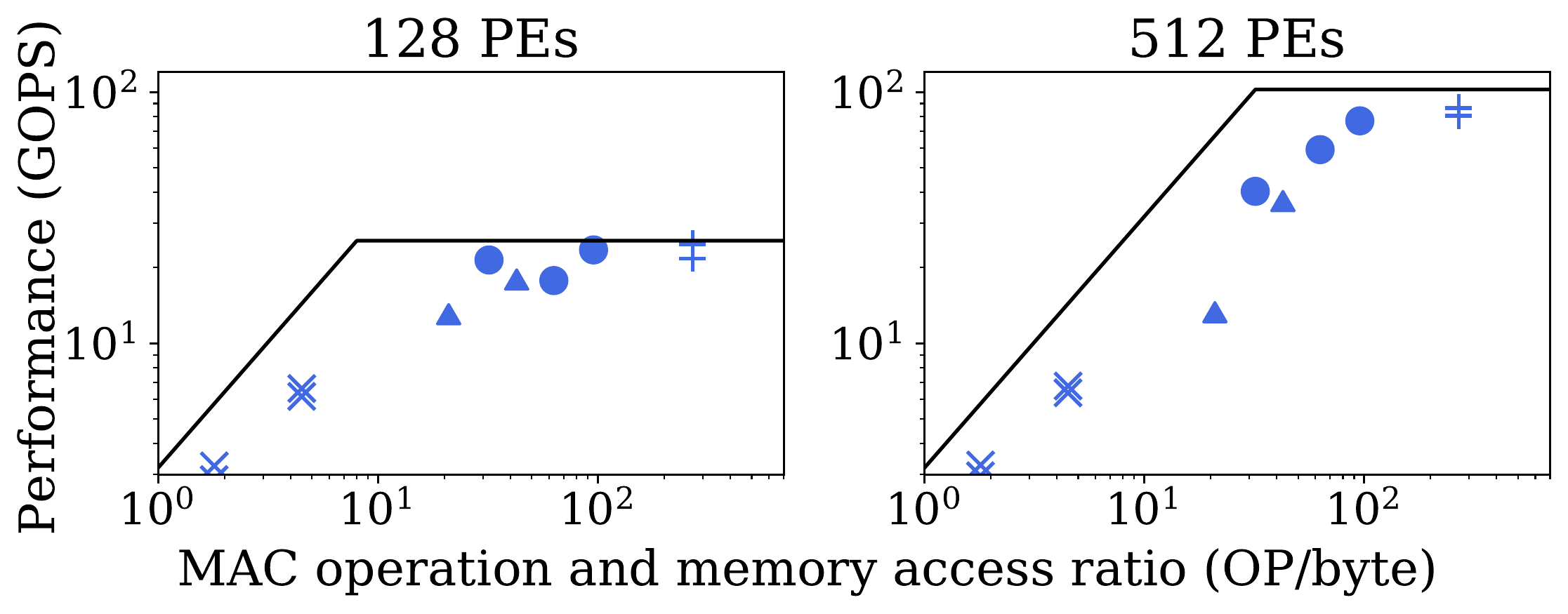}
\caption{{\bf Roofline analysis for workloads supported exclusively by \ArchName{}.} The performance results of \ArchName{} can reach the roofline for both memory-bounded and computation-bounded workloads.}\label{fig:exp:misc_workload}
\end{figure}

In \figref{exp:misc_workload}, the roofline analysis also shows that \ArchName{} can smoothly process layers with highly computation-bounded in recent networks like ESPCN~\cite{pixel_shuf} and DeepLab~\cite{deeplabv3}. For layers in MobileNet~\cite{mobilenet}, while the performance results are relatively low, we have already reached the roofline. This figure also demonstrates the spatial matching workloads at reasonably optimal performance.

%% file: large_float/table_networks.tex
\begin{table}[ht]
\centering
\footnotesize
\caption{Classic CNN workloads for benchmarking.}\label{tab:exp:standard_cnn}
\setlength{\tabcolsep}{1pt}
\begin{tabular}{|l|c|c|c|}
\hline
Layer & Stride & Kernel    & Channel  \\
      &   $s$  & $k_w,k_h$ & $C_i,C_o$\\
\hline
AL CONV1  & 4 & 11,11 & 3,48   \\
\hline
AL CONV2  & 1 & 5,5   & 48,128 \\
\hline
AL CONV3  & 1 & 3,3   & 128,192\\
\hline
AL CONV4  & 1 & 3,3   & 192,192\\
\hline
AL CONV5  & 1 & 3,3   & 192,128\\
\hline
TY CONV1  & 1 & 3,3   & 3,16   \\
\hline
TY CONV2  & 1 & 3,3   & 16,32  \\
\hline
TY CONV3  & 1 & 3,3   & 32,64  \\
\hline
\end{tabular}
\quad
\begin{tabular}{|l|c|c|c|}
\hline
Layer & Stride & Kernel    & Channel  \\
      &   $s$  & $k_w,k_h$ & $C_i,C_o$\\
\hline
TY CONV4  & 1 & 3,3 & 64,128 \\
\hline
TY CONV5  & 1 & 3,3 & 128,256\\
\hline
TY CONV6  & 1 & 3,3 & 256,512 \\
\hline
TY CONV8  & 1 & 1,1 & 1024,125 \\
\hline
IN $1\times7$ & 1 & 1,7 & 64,64   \\
\hline
IN $7\times1$ & 1 & 7,1 & 64,64   \\
\hline
SR CONV1  & 1 & 9,9 & 3,64    \\
\hline
&&&\\
\hline
\end{tabular}
\end{table}

%% file: large_float/table_all.tex
\begin{table}[ht]
\centering
\footnotesize
\setlength{\tabcolsep}{3pt}
\caption{The overall comparison of DNN architectures.}\label{tab:exp:perf_norm}
\begin{tabu}{l|rrr|rrr|}
\toprule
&\multicolumn{3}{c|}{128 PE}&\multicolumn{3}{c|}{512 PE}\\
\cline{2-7}
& TPU & Eyeriss & Vector- & TPU & Eyeriss & Vector-\\
&&& Mesh &&& Mesh\\
\midrule
Normalized GLB access  &   935 &    160 &     \textbf{42} &    534 &     55 &     \textbf{29}\\
Normalized DRAM access &   239 &     85 &     \textbf{45} &     71 &     \textbf{28} &     32\\
\hline
Area-efficiency ($P/AN$) & \textbf{22.55} &  12.48 &  20.49 &  15.91 &  11.12 &  \textbf{17.31}\\
\hspace*{1em}Performance ($P$, GOPS)  &    10 &     12 &     \textbf{20} &     27 &     41 &     \textbf{68}\\
\hspace*{1em}Area factor ($A$)        & 0.46 & 1.00 & 1.04 & 0.46 & 1.00 & 1.04\\
\hspace*{1em}Area multiplier ($N$)    & 1 & 1 & 1 & 4 & 4 & 4\\
\bottomrule
\end{tabu}
\end{table}

%% file: 50_conclusion.tex
\section{Conclusion}
In this paper, we proposed the VectorMesh architecture, discussed a workload scheduling process, and demonstrated its ability to execute various DNN and vision workloads closer to the performance roofline than other state-of-the-art architectures.  We provided practical implementations of the architecture and demonstrate its ability to reduce global buffer and DRAM fetches by 2-22 times and up to 5 times, respectively.

%% file: conference_101719.bbl
\begin{thebibliography}{10}

\bibitem{dram_power}
Mark Horowitz.
\newblock 1.1 computing's energy problem (and what we can do about it).
\newblock volume~57, pages 10--14, 02 2014.

\bibitem{scnn}
Angshuman Parashar, Minsoo Rhu, Anurag Mukkara, Antonio Puglielli, Rangharajan
  Venkatesan, Brucek Khailany, Joel Emer, Stephen W.~Keckler, and William
  Dally.
\newblock {SCNN}: An accelerator for compressed-sparse convolutional neural
  networks.
\newblock pages 27--40, 06 2017.

\bibitem{cnvlutin}
J.~{Albericio}, P.~{Judd}, T.~{Hetherington}, T.~{Aamodt}, N.~E. {Jerger}, and
  A.~{Moshovos}.
\newblock Cnvlutin: Ineffectual-neuron-free deep neural network computing.
\newblock In {\em 2016 ACM/IEEE 43rd Annual International Symposium on Computer
  Architecture (ISCA)}, pages 1--13, June 2016.

\bibitem{gratetile}
Y.~S. {Lin}, H.~C. {Lu}, Y.~B. {Tsao}, Y.~M. {Chih}, W.~C. {Chen}, and S.~Y.
  {Chien}.
\newblock Gratetile: Efficient sparse tensor tiling for cnn processing.
\newblock In {\em 2020 IEEE Workshop on Signal Processing Systems (SiPS)},
  pages 1--6, 2020.

\bibitem{tpu}
Norman~P. Jouppi et~al.
\newblock In-datacenter performance analysis of a tensor processing unit.
\newblock In {\em Proceedings of the 44th Annual International Symposium on
  Computer Architecture, {ISCA} 2017, Toronto, ON, Canada, June 24-28, 2017},
  pages 1--12, 2017.

\bibitem{eyeriss}
Y.~H. Chen, J.~Emer, and V.~Sze.
\newblock Eyeriss: A spatial architecture for energy-efficient dataflow for
  convolutional neural networks.
\newblock In {\em 2016 ACM/IEEE 43rd Annual International Symposium on Computer
  Architecture (ISCA)}, pages 367--379, June 2016.

\bibitem{roofline}
Samuel Williams, Andrew Waterman, and David Patterson.
\newblock Roofline: An insightful visual performance model for multicore
  architectures.
\newblock {\em Commun. ACM}, 52(4):65–76, April 2009.

\bibitem{systolic}
S.~Y. {Kung}.
\newblock {\em {VLSI array processors}}.
\newblock 1988.

\bibitem{tpu2}
Norman~P. Jouppi, Doe~Hyun Yoon, George Kurian, Sheng Li, Nishant Patil, James
  Laudon, Cliff Young, and David Patterson.
\newblock A domain-specific supercomputer for training deep neural networks.
\newblock {\em Commun. ACM}, 63(7):67–78, June 2020.

\bibitem{eyerissv2}
Yu{-}Hsin Chen, Joel~S. Emer, and Vivienne Sze.
\newblock Eyeriss v2: {A} flexible and high-performance accelerator for
  emerging deep neural networks.
\newblock {\em CoRR}, abs/1807.07928, 2018.

\bibitem{alexnet}
Alex Krizhevsky, Ilya Sutskever, and Geoffrey~E Hinton.
\newblock {ImageNet} classification with deep convolutional neural networks.
\newblock In F.~Pereira, C.~J.~C. Burges, L.~Bottou, and K.~Q. Weinberger,
  editors, {\em Advances in Neural Information Processing Systems 25}, pages
  1097--1105. Curran Associates, Inc., 2012.

\bibitem{tinyyolov1}
J.~{Redmon} and A.~{Farhadi}.
\newblock Yolo9000: Better, faster, stronger.
\newblock In {\em 2017 IEEE Conference on Computer Vision and Pattern
  Recognition (CVPR)}, pages 6517--6525, 2017.

\bibitem{srcnn}
C.~Dong, C.~C. Loy, K.~He, and X.~Tang.
\newblock Image super-resolution using deep convolutional networks.
\newblock {\em IEEE Transactions on Pattern Analysis and Machine Intelligence},
  38(2):295--307, Feb 2016.

\bibitem{pixel_shuf}
Wenzhe Shi, Jose Caballero, Ferenc Huszar, Johannes Totz, Andrew~P. Aitken, Rob
  Bishop, Daniel Rueckert, and Zehan Wang.
\newblock Real-time single image and video super-resolution using an efficient
  sub-pixel convolutional neural network.
\newblock In {\em The IEEE Conference on Computer Vision and Pattern
  Recognition (CVPR)}, June 2016.

\bibitem{dilconv}
Fisher Yu and Vladlen Koltun.
\newblock Multi-scale context aggregation by dilated convolutions.
\newblock In {\em ICLR}, 2016.

\bibitem{flownet}
A.~Dosovitskiy, P.~Fischer, E.~Ilg, P.~H{\"a}usser, C.~Haz{\i}rba{\c{s}},
  V.~Golkov, P.~v.d. Smagt, D.~Cremers, and T.~Brox.
\newblock {FlowNet}: Learning optical flow with convolutional networks.
\newblock In {\em IEEE International Conference on Computer Vision (ICCV)},
  2015.

\bibitem{eva}
Mark Buckler, Philip Bedoukian, Suren Jayasuriya, and Adrian Sampson.
\newblock {EVA}\({}^{\mbox{2}}\) : Exploiting temporal redundancy in live
  computer vision.
\newblock {\em CoRR}, abs/1803.06312, 2018.

\bibitem{costvol}
C.~Rhemann, A.~Hosni, M.~Bleyer, C.~Rother, and M.~Gelautz.
\newblock Fast cost-volume filtering for visual correspondence and beyond.
\newblock CVPR '11, pages 3017--3024, Washington, DC, USA, 2011. IEEE Computer
  Society.

\bibitem{systolic_bm1}
T.~{Komarek} and P.~{Pirsch}.
\newblock Array architectures for block matching algorithms.
\newblock {\em IEEE Transactions on Circuits and Systems}, 36(10):1301--1308,
  1989.

\bibitem{systolic_bm2}
L.~{De Vos} and M.~{Schobinger}.
\newblock {VLSI} architecture for a flexible block matching processor.
\newblock {\em IEEE Transactions on Circuits and Systems for Video Technology},
  5(5):417--428, 1995.

\bibitem{umi}
Y.~S. Lin, W.~C. Chen, and S.~Y. Chien.
\newblock Unrolled memory inner-products: An abstract gpu operator for
  efficient vision-related computations.
\newblock In {\em 2017 IEEE International Conference on Computer Vision
  (ICCV)}, pages 4587--4595, Oct 2017.

\bibitem{halide13}
Jonathan Ragan-Kelley, Connelly Barnes, Andrew Adams, Sylvain Paris, Fr{\'e}do
  Durand, and Saman Amarasinghe.
\newblock Halide: A language and compiler for optimizing parallelism, locality,
  and recomputation in image processing pipelines.
\newblock {\em SIGPLAN Not.}, 48(6):519--530, June 2013.

\bibitem{meritz}
Y.~{Lin}, W.~{Chen}, and S.~{Chien}.
\newblock {MERIT}: Tensor transform for memory-efficient vision processing on
  parallel architectures.
\newblock {\em IEEE Transactions on Very Large Scale Integration (VLSI)
  Systems}, pages 1--14, 2019.

\bibitem{maeri}
Hyoukjun Kwon, Ananda Samajdar, and Tushar Krishna.
\newblock {MAERI}: Enabling flexible dataflow mapping over dnn accelerators via
  reconfigurable interconnects.
\newblock In {\em Proceedings of the Twenty-Third International Conference on
  Architectural Support for Programming Languages and Operating Systems},
  ASPLOS '18, pages 461--475, New York, NY, USA, 2018. ACM.

\bibitem{gg3}
Hubert Nguyen.
\newblock {\em Gpu Gems 3}.
\newblock Addison-Wesley Professional, first edition, 2007.

\bibitem{inceptionv4}
Christian Szegedy, Sergey Ioffe, Vincent Vanhoucke, and Alexander Alemi.
\newblock Inception-v4, inception-resnet and the impact of residual connections
  on learning.
\newblock {\em AAAI Conference on Artificial Intelligence}, 02 2016.

\bibitem{deeplabv3}
Liang-Chieh Chen, George Papandreou, Florian Schroff, and Hartwig Adam.
\newblock Rethinking atrous convolution for semantic image segmentation.
\newblock 06 2017.

\bibitem{mobilenet}
Andrew~G. Howard, Menglong Zhu, Bo~Chen, Dmitry Kalenichenko, Weijun Wang,
  Tobias Weyand, Marco Andreetto, and Hartwig Adam.
\newblock {MobileNets}: Efficient convolutional neural networks for mobile
  vision applications.
\newblock {\em CoRR}, abs/1704.04861, 2017.

\bibitem{ramulator}
Y.~Kim, W.~Yang, and O.~Mutlu.
\newblock Ramulator: A fast and extensible {DRAM} simulator.
\newblock {\em IEEE Computer Architecture Letters}, 15(1):45--49, Jan 2016.

\end{thebibliography}
